\documentclass[sn-mathphys,iicol]{sn-jnl}

\jyear{2022}%

\theoremstyle{thmstyleone}%
\theoremstyle{thmstyletwo}%

\theoremstyle{thmstylethree}%

\begin{document}

\title[Humanities Citation Index (HuCI)]{The case for the Humanities Citation Index (HuCI): a citation index by the humanities, for the humanities}


\author*[1]{\fnm{Giovanni} \sur{Colavizza}}\email{g.colavizza@uva.nl}
\equalcont{All authors contributed equally to this work.}

\author[2,3]{\fnm{Silvio} \sur{Peroni}}\email{silvio.peroni@unibo.it}
\equalcont{All authors contributed equally to this work.}

\author[4]{\fnm{Matteo} \sur{Romanello}}\email{matteo.romanello@unil.ch}
\equalcont{All authors contributed equally to this work.}

\affil*[1]{\orgdiv{Institute for Logic, Language and Computation (ILLC)}, \orgname{University of Amsterdam}, \orgaddress{\street{Science Park 107}, \city{Amsterdam}, \postcode{1098 XG}, \country{The Netherlands}}}

\affil[2]{\orgdiv{Research Centre for Open Scholarly Metadata, Department of Classical Philology and Italian Studies}, \orgname{University of Bologna}, \orgaddress{\street{via Zamboni 32}, \city{Bologna}, \postcode{40126}, \country{Italy}}}

\affil[3]{\orgdiv{Digital Humanities Advanced Research Centre (/DH.arc), Department of Classical Philology and Italian Studies}, \orgname{University of Bologna}, \orgaddress{\street{via Zamboni 32}, \city{Bologna}, \postcode{40126}, \country{Italy}}}

\affil[4]{\orgdiv{Institute of Classical Studies and Archaeology}, \orgname{University of Lausanne}, \orgaddress{
\city{Lausanne}, \postcode{1015}, \state{Switzerland}}}


\abstract{Citation indexes are by now part of the research infrastructure in use by most scientists: a necessary tool in order to cope with the increasing amounts of scientific literature being published. Commercial citation indexes are designed for the sciences and have uneven coverage and unsatisfactory characteristics for humanities scholars, while no comprehensive citation index is published by a public organization. We argue that an open citation index for the humanities is desirable, for four reasons: it would greatly improve and accelerate the retrieval of sources, it would offer a way to interlink collections across repositories (such as archives and libraries), it would foster the adoption of metadata standards and best practices by all stakeholders (including publishers) and it would contribute research data to fields such as bibliometrics and science studies. We also suggest that the citation index should be informed by a set of requirements relevant to the humanities. We discuss four: source coverage must be comprehensive, including books and citations to primary sources; there needs to be chronological depth, as scholarship in the humanities remains relevant over time; the index should be collection-driven, leveraging the accumulated thematic collections of specialized research libraries; and it should be rich in context in order to allow for the qualification of each citation, for example by providing citation excerpts. We detail the fit-for-purpose research infrastructure which can make the humanities citation index a reality. Ultimately, we argue that a citation index for the humanities can be created by humanists, via a collaborative, distributed and open effort.}

\keywords{citation indexing, open citations, bibliographic information retrieval, open research infrastructure}



\maketitle

\section{Introduction}\label{sec:introduction}



Citation  indexes  are  by  now  part  of  the  research  infrastructure  in  use  by  most  scientists:  a  necessary  tool  in  order  to  cope  with  the  increasing  amounts  of  scientific  literature  being  published. However, existing commercial  citation  indexes  are  designed  for  the  sciences  and  have  uneven  coverage  and  unsatisfactory  characteristics  for humanities\footnote{Throughout this paper we use the term \textit{humanities} as a shorthand for Arts \& Humanities (A\&H). To a degree, the Social Sciences are also concerned.} scholars. This situation has both discouraged the usage of citation indexes and hindered bibliometric studies of humanities disciplines.  

The creation of a citation index for the humanities may well appear as a daunting task due to several characteristics of this field, such as its fragmentation into several sub-disciplines, the common practice of publishing research in languages other than English, as well as the amount of scholarship from past centuries that is still waiting to be digitised.

Notwithstanding these challenges, we argue that the creation of such an index can be highly beneficial to humanities scholars for, at least, the following reasons. Firstly, humanities scholars have long been relying on information seeking behaviours that leverage citations and references lists for the discovery of relevant publications -- a strategy that citation indexes are designed to support and facilitate. Secondly, a comprehensive citation index for the humanities will be a valuable source of data for researchers willing to conduct bibliometric studies of the humanities. Lastly, capturing the wealth of references to primary and secondary sources contained in humanities literature will allow to create links between archives, galleries, libraries and museums where digitized copies of these sources can increasingly be found.

Before continuing with this paper, we introduce key terminology related to citation indexing that will be used throughout this paper, adopting the definitions from \cite{peroni_opencitations_2018}. These are: bibliographic entity, bibliographic resource and bibliographic citation.
A \textbf{bibliographic entity} is any entity which can be part of the bibliographic metadata of a bibliographic artifact: it can be a person, an article, an identifier for a particular entity (e.g., a DOI), a particular role held by a person (e.g., being an author) in the context of defining another entity (e.g., a journal article), and so forth. A \textbf{bibliographic resource} is a kind of bibliographic entity that can cite or be cited by other bibliographic resources (e.g., a journal article), or that contains other  resources (e.g., a journal). A \textbf{bibliographic citation} is another kind of bibliographic entity: a conceptual directional link from a citing bibliographic resource to a cited bibliographic resource.
The citation data defining a particular citation must include the representation of the conceptual directional link of the citation and the basic metadata of the involved bibliographic resources, that is to say sufficient information to create or retrieve textual bibliographic references for each of the bibliographic resources. Following \cite{peroni_open_2018}, we say that a bibliographic citation is an open citation when the citation data needed to define it are compliant with the following principles: structured, separate, open, identifiable, available.

The remaining of this paper is organised as follows. In Section \ref{sec:related-work} we discuss previous work on analysing the behaviour of humanities scholars in relation to information retrieval. We also present the main limitations of existing citation indexes, seen from the perspective of the humanities, and outline the main obstacle that citation indexing has faced in this area. In Section \ref{sec:citation-index-AH-needs} we argue for the need of a Humanities Citation Index (HuCI from now onward) and in Section \ref{sec:citation-index-characteristics} we present what we believe are the essential characteristics that such an index should have. We then propose a possible implementation of HuCI, based on a federated and distributed research infrastructure (Section \ref{sec:research-infrastructure}). We conclude with some considerations on how HuCI relates to recent efforts to create open infrastructures for research.
\section{Related Work}\label{sec:related-work}

\subsection{On scholarly information retrieval in the A\&H}\label{sec:scholarly-IR}

The needs and behaviours of humanities scholars in terms of information seeking has been an active area of study especially in the field of Library and Information Science (LIS), where research on this topic started in the 1980s and early 1990s \cite{stone_humanities_1982, ellis_behavioural_1989, watson-boone_information_1994}. For a thorough review of the early literature on this topic see \cite{wiberley_jr_humanities_2009}[p. 2198] and \cite{benardou_understanding_2010}[pp. 19-21].
Determining the information needs and behaviours of humanities scholars was essential for librarians in order to support scholars in their research by devising new library systems or by improving the guidelines for abstracting publications to cater for the specific needs of humanities
scholars \cite{tibbo_abstracting_1993}. What emerges from this literature are also the key strategies for finding bibliographic information that characterises humanities scholarship. Firstly, scholars use proper names extensively when searching as compared with scholars in other disciplines \cite{wiberley_patterns_1989, bates_getty_1996, palmer_scholarly_2009}. Secondly, a prominent behaviour among humanities scholars is to search for bibliographic information by browsing \cite{bates_getty_1996, ellis_behavioural_1989, meho_modeling_2003}. A typical example is browsing books in the stacks or shelves of a library. What characterises browsing as opposed to a targeted search is that it favours the serendipitous discovery of relevant information: the physical proximity of books on library shelves, which is related to their subject classification, may in some cases transcend the boundaries of subjects. Finally, a third prominent search strategy is the already mentioned citation chaining with its two variants of backward and forward chaining \cite{ellis_behavioural_1989,buchanan_information_2005}. The former consists of starting from one publication – the seed document -- and then following up the references it contains in order to expand the initial search and to discover other related publications. The latter consists of starting from a seed document and then finding which other publications cite it. Moreover, an empirical study of the information seeking strategies of humanities scholars reports that searching and browsing proved to be rather ineffective strategies for locating information and that citation chaining was the most common behavioural pattern \cite{buchanan_information_2005}[pp. 227–228].

\subsection{Citation indexing and the Humanities}\label{sec:citation-indexing}

Citation indexing is commonplace for Science, Technology, Engineering and Mathematics (STEM) literature. Mainstream indexes such as Google Scholar, the Web of Science, Scopus, Dimensions or Semantic Scholar are largely capable of indexing most citations accurately. To be sure, their coverage is still uneven and far from uniform \cite{martin-martin_google_2021, visser_large-scale_2021}. One of the critical problems which are left open is the uneven coverage of different disciplines, with those part of the arts and humanities usually faring worse than most \cite{harzing_google_2016}. Several reasons for this state of affairs have been individuated, which can be grouped into two categories. Intrinsic factors, which depend on the literature itself, and extrinsic factors, which depend on the information environment where citation mining is performed \cite{colavizza_citation_2019-1}. 

\textit{Intrinsic factors} which act as obstacles to citation indexing in the humanities include the more limited availability of born digital or digitized publications, a higher variety of languages and publication venues in use, the practice to publish monographs, complex referencing practices and motivations which limit their automatic processing. These topics have been amply discussed in the literature \cite{kulczycki_publication_2018, hicks_difficulty_1999, nederhof_bibliometric_2006, huang_characteristics_2008, santos_citing_2021}. \textit{Extrinsic factors} have been less the focus of previous work and include, instead, the variety and fragmentation of catalogs, information systems and other sources of unique identifiers and authoritative metadata. These issues are well-known more generally in the Galleries, Libraries, Archives, and Museums (GLAM) sector. A recent study on metadata aggregation highlights several characteristics of this landscape, among which these fall within what we here refer as extrinsic factors \cite{freire_cultural_2020}:
\begin{itemize}
\item Each GLAM sub-domain (libraries, archives and museums) applies its specific resource description practices and data models.
\item All sub-domains embrace the adoption and definition of standards-based solutions addressing description of resources, but to different extents.
\item Interoperability of systems and data is scarce across sub-domains, but it is somewhat more common within each sub-domain, at the national and the international levels.
\end{itemize}

As a consequence of the limitations enacted by both intrinsic and extrinsic factors, it is more difficult to comprehensively index the humanities via citations, a condition that limited the use of quantitative bibliometric methods in this area \cite{ardanuy_sixty_2013}, despite clear progress over recent time \cite{petr_journal_2021, hammarfelt_beyond_2016}. The lack of a comprehensive and reliable citation index remains a known and open problem in the humanities \cite{heinzkill_characteristics_1980, linmans_why_2009, sula_citations_2014}. Our contribution proposes a way forward which mainly addresses the obstacles posed by extrinsic factors, and is true to the way the humanities communicate research and retrieve scholarly information.
\section{The need for a citation index for the humanities}\label{sec:citation-index-AH-needs}

Scholarship in the humanities rests on solid traditions, most crucially developed in the archives, libraries and information studies communities. It is thus worth asking the question: \textbf{why do we need a citation index for the humanities?} We advance four motivations: to dramatically improve current scholars’ information retrieval capabilities; to interlink presently siloed GLAM information systems; to foster best practices in terms of referencing and metadata; to provide for research data to bibliometrics and science studies.

\subsection{Improve scholarly information retrieval}

From an information retrieval point of view, citation indexes seem to be the natural evolution of disciplinary and thematic bibliographies (e.g., the \textit{Annual Bibliography of English Language and Literature}\footnote{\url{https://www.mla.org/Publications/MLA-International-Bibliography}.} or \textit{L'Année Philologique}\footnote{\url{https://about.brepolis.net/lannee-philologique-aph}.}), which are widely used by scholars across the Humanities to conduct literature search. A citation index, in fact, can be seen a bibliography whose entries are linked with one another depending on the citations that are found in the full-text of the catalogued publications. Moreover, thematic bibliographies such as the \textit{World Shakespeare bibliography}\footnote{\url{https://www.worldshakesbib.org}.} or the \textit{International Dante Bibliography}\footnote{\url{[https://bibliografia.dantesca.it](https://bibliografia.dantesca.it)}.} often provide users with the ability to search for publications related to specific literary works -- a functionality that could also be provided by a citation index which captures references to primary sources.

Despite the existence of bibliographies and bibliographic databases, Humanities scholars cannot yet fully rely on citation indexes when searching for secondary literature, nor to keep up to date with recent developments (e.g., via citation alerts). As we highlighted above, it is the limited coverage of existing citation indexes more than any intrinsic limitation that has been the decisive factor in discouraging their more systematic adoption in retrieval practices. This need not be a sealed fate. Assuming sufficient coverage, in both quality and quantity, a citation index for the humanities can first and foremost serve the same information retrieval needs these tools provide for in the sciences since decades. It is likely that a non-negligible fraction of humanities scholars already uses services such as Google Scholar and Google Books \cite{chen_exploring_2019}, even in the absence of comprehensive evidence on their coverage and reliability.
 
Furthermore, a variable yet non-negligible amount of references in the humanities are given to primary sources, such as archival documents or literary works \cite{knievel_citation_2005}. There has never been a way to count and retrieve all references to a given primary source without painstaking manual work. Knowledge about primary sources, in terms of their existence, location and means of access, takes up a substantial amount of time and training in the humanities, sometimes becoming all too treasured. In principle, both primary and secondary sources should be indexed in the humanities citation index. This will allow anyone to immediately gauge which sources have been used together, where and by whom. In practice, several open challenges will need to be overcome first, including programmatic access to uniform GLAM metadata.

\subsection{Interlink GLAM collections via citations}\label{sec:interlinking-collections-via-citations}

GLAM information ecosystems often exist in isolated silos: metadata and data are largely made accessible by the specific institution that creates and curates them. Notable exceptions exist, for example national library catalogs and projects such as Europeana. Nevertheless, to the best of our knowledge, no encompassing information retrieval infrastructure exists spanning across GLAM institutional categories, for example interlinking libraries (L) with archives (A). Citation links extracted from scholarly literature can do just that.
 
The literature in the humanities in fact contains a wealth of references to primary sources, accumulated over centuries of scholarly work. Within the scope of one project alone, some of the authors were able to extract nearly 700,000 references to primary sources from approximately 1900 books and 5500 journal articles (Venice Scholar Index\footnote{\url{https://venicescholar.dhlab.epfl.ch}} \cite{colavizza_linked_2018}). Citation links connect secondary literature, hence library catalogs, with archives, galleries and museums’ information systems. They also connect archives, galleries and museums directly by virtue of co-citation relationships (i.e., two resources are connected if they are cited together by a third one). These links effectively constitute a dormant virtual information system which awaits to be digitally materialized. By so doing, a significant acceleration and democratization to the access of primary sources can be realized, contributing to a broader scholarly and public engagement with these collections as is currently the case.
 
Digitally materializing citation links would create incentives to make GLAM information retrieval and research infrastructure increasingly more interoperable and interdependent, to the great benefit of the research community. Citation indexing requires publication data and metadata, which must be made available by publishers and GLAM institutions. We argue that, once the benefits of citation indexing will have been made tangible to a sufficient degree, this will create a positive feedback loop for all stakeholders to gradually improve on their practices in order to make citation indexing increasingly easier and automatic.

\subsection{Improve current practices}
The automatic extraction and indexing of structured information, such as bibliographic citations, typically requires a high degree of openness and standardization in the ecosystem it happens in. Citation indexing requires open, standardized and programmatically accessible metadata about primary and secondary sources alike, as well as access to the full text of scholarly publications. It also benefits from a high degree of uniformity in the referencing practices of authors, which makes reference parsing all the more feasible. Yet, all this is costly, hard and time consuming. For all stakeholders to strive to higher openness, standardization and accessibility, we require a positive incentive. We argue that citation indexing, once it reaches a certain threshold, actually provides for one: if a community starts using citation indexes for information retrieval, being indexed increasingly becomes a necessity, hence related investments will be made.
 
Citation indexing starts with authors. Referencing practices, sometimes less than uniform and coherent, pose a significant challenge to the automatic extraction of citations (e.g., \cite{nederhof_bibliometric_2006}). Yet, once references become data, and their value as links is immediately made tangible via citation indexing, authors might have more incentives to make their referencing practices syntactically and stylistically more uniform in view of improving their harvesting and correct indexing.
 
A similar point in case can be made for publishers. On the one hand, proof-checking work can make sure to provide for uniform references with sufficient information for their indexing, similarly to what is provided by several scientific publishers. On the other hand, and more importantly, publishers could sign up (and effectively contribute) to the Crossref and OpenCitations initiatives, making their metadata and citation data available. The existence of a citation index for the humanities should foster participation in such initiatives. Failing that, or considering the backlog of already published publications (especially if printed), GLAM institutions themselves can take a leading role, as we discuss below.
 
The positive incentive to expose open, standardized and programmatically accessible metadata provided by the citation index will also apply to GLAM institutions, once the benefits of interlinked collections and increased visibility/searchability will become apparent. A crucial challenge for us will be to reach a critical mass of citation data to provide for an indispensable service to a sizable share of the research community and, at the same time, initiate the positive incentive for all stakeholders.

\subsection{Research data for bibliometrics and science studies}
 
It is well known that the humanities are significantly understudied by the bibliometrics and quantitative science studies community, largely because of the lack of citation data \cite{colavizza_citation_2019}. This has several consequences, among which the separation of qualitative studies on the humanities from analyses grounded in (bigger) data \cite{franssen_science_2019}. Furthermore, it also causes a widespread science-as-the-norm/humanities-as-an-exception mindset in bibliometrics and research evaluation as a whole, as if it were the case that citations cannot be used to study the humanities. To be sure, indexing citations in the humanities is challenging, yet it would allow the bibliometrics and quantitative science studies communities to finally approach the humanities on equal ground with respect to the sciences.
The proposed citation index for the humanities can radically alter this state of affairs. First of all, a bibliometrics for the humanities grounded in data as well as theory could finally be developed, in full recognition of the specificities of the humanities \cite{hammarfelt_beyond_2016}. Secondly, citation data in the humanities is very rich, if we consider the varied publication typologies, languages, primary and secondary sources that come into play. As a consequence, citation data from the humanities will require novel methods and approaches that might not only provide insights into these data, but as well inform further developments when applied to citation data from the sciences. The HuCI can essentially put an end to the age of the so-called ``non-bibliometric'' humanities.

Citation data, once available, have been used for research evaluation. Indicators such as citation counts or the H-index, are widespread and have been amply discussed by the bibliometrics community \cite{waltman_review_2016}. Recently, public efforts have been made to call for a redress and improvement in the use of citation-based indicators \cite{hicks_bibliometrics_2015}.\footnote{Also see the San Francisco Declaration on Research Assessment (DORA): \url{https://sfdora.org/read}.} It will be likely unavoidable to face similar discussions if and when the HuCI materializes. We believe these worries should not prevent it from happening, for the very reasons we just detailed. Furthermore, HuCI could provide for an opportunity to rethink the way we use citation-based indicators in research evaluation. The humanities have a long-lasting tradition of peer review assessment which, when mixed with situated and contextualized metrics (which in turn need not be just citation-based), has the potential to inform research evaluation in the  sciences too.
\section{The characteristics of a citation index for the A\&H}\label{sec:citation-index-characteristics}

Having clarified why we believe a citation index for the humanities is motivated, we detail here four requirements we propose it should have. These are: comprehensive source coverage and chronological depth, rich information provided to contextualize citations, and a growth strategy driven by institutional collections. We note that we intend these requirements as something to aspire to: they represent end goals more than necessary conditions to begin with.

\subsection{Source coverage}
Scholars in the humanities use a complementary variety of publication typologies, such as monographs, journal articles and contributions in edited volumes. Journal articles, the main focus of existing commercial citation indexes, in general account for a small fraction of the output in all the humanities \cite{knievel_citation_2005}. We thus argue that the first requirement of a citation index for the humanities is complete coverage in terms of publication typologies. A related requirement, or pain point, is multilingualism. Scholarly literature in national languages abounds in most of the humanities, yet this variety is not often captured by digital resources. A case in point is the situation of Classics: 75\% of Classics publications contained in JSTOR are written in English, while the language of publications reviewed in L’Année Philologique (APh, the most important bibliography in this field) is much more evenly distributed between English, German, Italian and French. In fact, Scheidel \cite{scheidel_continuity_1997} reports that, of the publications reviewed by APh in 1992, 30\% were written in English, roughly 25\% in Italian, 20\% in French and 20\% in German. Ideally, language should not be a source of bias in the citation index.\footnote{Promoting measures against language bias in the context of research assessment is one of the three key recommendations made by the Helsinki Initiative \cite{federation_of_finnish_learned_societies_helsinki_2019}.}
 
As we anticipated above, the second requirement we put forth is the full indexation of citations to primary sources. Interestingly, this requirement compels a discussion of citation granularity: what is the object of a reference which should be considered in a citation index? Typically, for secondary literature we use the level of the work in FRBR terms \cite{ifla_functional_1997}. Hence, citations are accumulated for, say, a journal article aggregating over all its expressions (e.g., in pre-print and printed versions) or for a book over all its editions, excluding those with major revisions that justify calling it a new work. For primary sources, we typically consider unique items (the lowest FRBR level), for example archival documents or unique artworks. We consider instead works when, say, dealing with critical editions of a classic author, where the editing activity is considered scholarly and the source is printed into editions. All this to say that the choice of the citation aggregation object is far from straightforward for primary sources, and a good rule of thumb is that further aggregation is always possible, while disaggregation can be more difficult to undo. Hence, we recommend lower FRBR citation aggregation levels when in doubt.

\subsection{Chronological depth}
The humanities are known to publish at a relatively slower pace than other sciences and to keep citing older relevant literature (e.g., \cite{nederhof_bibliometric_2006, hellqvist_referencing_2010}). This has two consequences for the citation index: first, and foremost, it is crucial to index older literature as well, spanning back ideally to when systematic scholarly referencing became commonplace \cite{grafton_footnote_1999}. Secondly, and this is not a requirement but an opportunity we highlight, digitizing and making openly available old and out-of-copyright literature, in conjunction with its indexation via citations, would constitute a great service to scholars. It would not only improve the use of such literature, but open up opportunities to study the history of scholarship in the humanities at unprecedented scale and comprehensiveness.

\subsection{Rich in context}
Previous work has elucidated how the citation semantics in the humanities tend to be rich and varied \cite{hellqvist_referencing_2010}. This is of crucial importance when using citations for information retrieval: is a citation supportive or dismissing? Is it contextual, perfunctory or does it substantially underpin an argument? The citation index we propose will need to make every effort possible to offer its users all the means necessary to appreciate and understand every citation link. This is mainly done by providing relevant context, within the bounds of existing copyrights.
 
Citation contexts are the excerpts of text preceding and following a citation. The most common context, in this sense, is the sentence where a citation is made. Nevertheless, a context can cover any relevant span, e.g., a few sentences or a whole paragraph. Another source of contextual information is given by proximal co-citations: which other sources are cited with the one under consideration, within the same citing publication? Lastly, providing the exact details of the citation, such as the page number it refers to, also helps to specify its scope. It is possible to see citations and their contexts in aggregate, from the point of view of either the citing or cited sources. This is the case when we consider, for example, all the other sources a given source is co-cited with. It is also possible to consider every citation as situated in a quite specific location of a publication. For example, by considering co-cited sources within the same paragraph of a well-defined citing publication. Both views, the aggregate and the detail, provide for relevant contextual information for a scholar to interpret citation links, and use them for information retrieval.

\subsection{Collection driven}
We conclude this section not by discussing a requirement, but by suggesting a growth strategy for HuCI. Mainstream citation indexes convey the impression, and sometimes the illusion, of comprehensive coverage. Only when we are able to trust a citation index in this sense, we, as scholars, can rely on it for our work. If a citation index is manifestly incomplete, and especially if what is missing is unknown or hard to qualify, it will be difficult for it to succeed. Given the daunting task we have set ourselves to with the humanities citation index and the stated requirements, we also need a reasonable growth strategy. Our proposal is to be topic/collection driven. That is to say, we recommend to index topically coherent batches of scholarly literature, by leveraging the specialized collections of research libraries.
 
In our previous work on the historiography of Venice, we faced the task of defining the limits of what pertains to this topic and what can be left outside. By relying on a set of finding aids – library catalogs, bibliographies, shelving strategies and specialized collections – we were able to create a coherent citation corpus \cite{colavizza_references_2018}. We suggest here that this approach can make HuCI scale, one topic/collection at the time. In so doing, the citation index can gradually serve more and more and larger and larger humanities communities.

Creating the humanities citation index requires not only a growth strategy, but first and foremost a research infrastructure which provides for the right affordances to build the index as a collaborative, distributed and open effort. We propose its design in what follows.

\subsection{Metadata ecosystem and requirements}\label{sec:metadata-ecosystem-requirements}

In the research infrastructure needed to build an A\&H citation index, libraries play a key role not only as holders of digitized collections but also as potential providers of data that can greatly support the citation extraction process. In fact, library catalogues constitute highly valuable knowledge bases of bibliographic information that can be exploited when doing citation mining, and especially citation matching. 

We identify a set of key technical requirements that need to be met if library catalogue metadata are to be seamlessly integrated into the HuCI infrastructure. These requirements are:
\begin{enumerate}
    \item ability to handle the heterogeneity of metadata formats;
    \item provision of unique persistent identifiers;
    \item machine-aided creation, delivery and exchange of metadata;
    \item fine-grained/granular metadata descriptions;  
    \item open licensing of metadata.
\end{enumerate}

\textbf{Metadata formats}. From the point of view of citation mining pipelines and processes there is a need to have metadata expressed in concise and ``easy-to-process'' formats. Such concerns become even more relevant when the metadata processing happens at a large scale, as the needs arise for optimising processing time and for efficient data storage. For example, in the context of previous work carried out by some of the authors \cite{colavizza_linked_2018,colavizza_citation_2019}, the Central Institute for the Union Catalogue of Italian Libraries and for Bibliographic Information (ICCU) has created a dump of 15 million records by transforming its data from MARC to a JSON-based representation, so as to facilitate their use in the project’s citation mining pipeline. Along similar lines, \cite{bergamin_new_2018} have successfully tested a workflow for mapping ICCU’s UNIMARC data onto Wikibase Data Model, which would allow for using Wikibase as an environment to manage and edit bibliographic data, as well as exposing such data in an easier to process format.  

MARC, however, is only one of the many formats that characterise the landscape of library metadata, where a plethora of old and new formats co-exist \cite{tennant_bibliographic_2004}. This situation makes it seem rather unlikely that libraries will converge to a common and widely-adopted metadata format in the near future. As a result, a key requirement of the HuCI infrastructure is the ability to handle this heterogeneity of bibliographic metadata formats, achievable by developing code modules that read these formats and map them onto a common one. 

\textbf{Provision of unique identifiers}. In addition to the granularity of descriptions, the provision of unique, persistent identifiers to identify bibliographic resources is another key requirement for metadata that are meant to support citation mining processes. Ideally, any primary or secondary source of which we are interested in tracking the citations ought to be identifiable by means of a unique, persistent identifier (e.g., a resolvable URI). Naturally, what is considered as a primary source varies from domain to domain: archival documents in History, various types of texts in Classics (e.g., canonical, papyri, inscriptions), manuscripts in Medieval Literature Studies, inscriptions and papyri in Egyptology, and so forth. Once these identifiers are in place, it is possible to use them to link `disambiguated' citations. However, it cannot be the task of a single project to mint and provide these identifiers. This process should be happening in each discipline -- and it has already been happening over the past years e.g., in Classics \cite{romanello_using_2017} -- but it can be fostered and accelerated by large-scale initiatives involving libraries and cultural heritage institutions, such as the European Open Science Cloud \cite{hellstrom_second_2020}.

\textbf{Machine-aided creation, delivery and exchange of metadata.} There is an urgent need to take humans \textit{out} of the loop insofar as access to and exchange of library metadata are concerned. Libraries -- and especially aggregators of library metadata (e.g., national aggregators, library consortia, etc.) -- ought to provide, at the very least, regular data dumps of their bibliographic metadata so as to facilitate their consumption and further reuse. Data dumps, however, being frozen snapshots of a dataset, raise the issue of synchronisation between the data at the source and the copy of the data used by other systems and processes. A partial solution to this problem is to provide streams of data (e.g., via APIs) in addition to regular dumps. 

\textbf{Granularity of bibliographic descriptions.} The granularity of bibliographic descriptions is an apt example of gaps currently existing between the needs and requirements of citation mining projects, on the one hand, and the cataloguing practices currently adopted by the majority of libraries, on the other hand. Types of publications where granularity matters the most are journal articles, book chapters and individual essays within collective volumes. In fact, while the citation unit of such publications is often the most granular (e.g., a given journal article, as opposed to the entire journal), cataloguing practices often do not reach that level of granularity in bibliographic descriptions. 

\textbf{Open licensing.} Despite a declared willingness to share, often libraries and other cultural heritage institutions make available online data dumps that do not come with explicitly defined (open) licenses. They ought to be encouraged to always provide explicit license statements, as their absence hinders the reuse of shared data by others.
\section{Research infrastructure}\label{sec:research-infrastructure}

\begin{figure}[h]%
\centering
\includegraphics[width=0.4\textwidth]{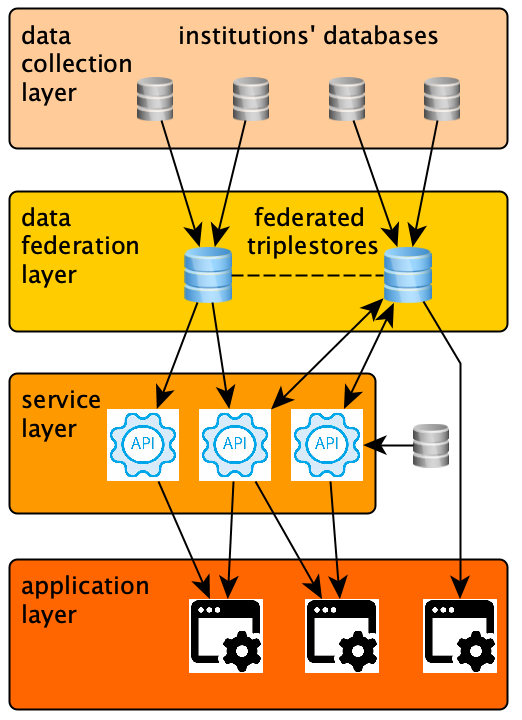}
\caption{The 4-layer architecture of the Humanities Citation Index. The solid arrows show that an item in one layer (ending node) uses the information provided by an item in another layer (starting node). The dashed line highlights existence of a federation mechanism between the items of the same layer.
\label{fig:huci-high-level}}
\end{figure}

We propose to adopt a federated and distributed approach to design the research infrastructure required to create the Humanities Citation Index. Such an approach implies that the creation of citation data is delegated to a federation of cooperating institutions rather than being carried out by a single, central entity. The scenario we envisage (see Fig. \ref{fig:huci-high-level}) is having a network of GLAM institutions, each of them contributing citation data extracted from their digitised collections through a common open source software platform. These data will then be harvested, aggregated and consolidated to become the HuCI citation corpus, available to researchers both via search and exploration interfaces, as well as data dumps to be further analysed and visualised through external tools. In what follows we describe in more detail the proposed architecture, as well as the challenges related to its implementation.

\subsection{Distributed and federated approach}

A distributed and federated approach recognizes the central role that libraries and other cultural heritage institutions could play with respect to the curation of their digitized collections. A modern notion of collection curation, we argue, ought to include the extraction of structured contents (e.g., citations) from digitized materials. This could take the form of manual verification, carried out by librarians, of automatically extracted information, as advocated by \cite{lauscher_linked_2018} for the specific case of citations.   

At a technical level, a federated model has the advantage that it gives individual institutions a certain degree of freedom in deciding what can and should be made openly accessible -- thus harvested from a federation of partner institutions -- and what, on the contrary, should remain accessible only internally, within the boundaries of institutional access. We can call this access model ``open by default and closed by necessity''. A typical case where it proves useful is the display of contextual information about citations (i.e., an excerpt of the text surrounding the citation), which can be complicated due to copyright restrictions. A library holding digitized materials under copyright will want to give exclusively to its users full access to citation contexts from these publications, while still sharing with the wider community data extracted from these publications that do not fall under copyright, such as citation data. It is worth mentioning on this respect that citation data are just facts, and as such cannot be copyrighted. Thus, following the guidelines in \cite{peroni_open_2018} and the Initiative for Open Citations (I4OC)\footnote{\url{https://i4oc.org}.}, they should be released as public domain material using appropriate solutions, such as the CC0 waiver.

Moreover, a distributed approach makes sense not only for the collection of citation data but also for the hosting of the resulting citation corpus. In fact, collecting all the citation data from the whole scientific knowledge in one single, centralized repository -- albeit feasible -- would raise considerable issues in terms of maintenance and performances. For instance, consider that the data available in the OpenCitations Corpus includes over 7.5 million bibliographic resources, mainly (~80\%) journal articles and their issues, volumes, and journals \cite{damato_one_2017}. Their bibliographic metadata and their provenance information occupy more than 2 billion triples. Supposing that, roughly, 45 million new journal articles are published every year \cite{van_harmelen_end_2017} and considering that 10\% belong to the A\&H, we can estimate 4.5 million new articles in the A\&H every year. This amount of data can be stored in 230 gigabytes using the model adopted by OpenCitations -- thus, we will need more than 5.7 terabytes for storing the metadata of all the A\&H articles published in the ``Web-era'', since 1994. While such amount of bytes can be even manageable in a big file system, these figures can drastically increase if we want to keep track of all the citation links among articles, if we start to ingest data coming from books (that are the primary publication object for the A\&H which contains more citation links than any other scholarly medium), and if we consider publications that are older than 25 years. 

The availability and cost of storage is something that will become more manageable in time. However, there is another, probably more pressing issue concerning the scenario mentioned above: having an infrastructure that guarantees efficient data querying is something very demanding when a large amount of data is actually available in a single and centralised database. As rough estimate considering the figures above, we would require 30 billions triples for handling the 4.5 million articles in A\&H in the past 25 years. It is worth mentioning that this figure does not include books and all the older literature, nor any further extension of the kinds of data to store – for instance, the figures above are based on what is currently stored in the OpenCitations Corpus and, as such, do not account for abstracts or authors’ affiliations. Therefore it emerges that a centralised solution like storing all the data in a single database, is not feasible in the long-term.

\subsection{Architecture}

The HuCI infrastructure that we propose consists of four main layers, whose interplay is schematically depicted in Fig. \ref{fig:huci-high-level}:
\begin{enumerate}
    \item The \textbf{data collection layer} consists of a network of content providers, who hold digitised materials (be they public or private) and contribute to the growth and coverage of the citation index by making openly available the citation data extracted from their holdings.
    \item The \textbf{data federation layer} is conceived as a federation of decentralised citation databases based on RDF technologies where the HuCI citation data is actually stored.
    \item The \textbf{service layer} provides HTTP APIs that allow for standardised access to HuCI data (e.g., via SPARQL endpoints and common REST Web APIs), external resources (e.g., those included in library or archive metadata) and services (e.g., author disambiguation).
    \item Finally, the \textbf{application layer} is an ecosystem of tools and software components, plugged on HuCI's virtual triplestore (via the previous layer or by consuming directly the data from the data federation layer), that allow A\&H researchers to discover and identify relevant literature for their research, and provides bibliometric insights into the citation data.
\end{enumerate}

In what follows we discuss in greater detail each of these infrastructure components. In addition to such components, it is important that the various providers of citation data are compliant as much as possible with the Principles of Open Scholarly Infrastructures (POSI) introduced in \cite{bilder_principles_2015}. These principles are organised in three themes: Governance, Sustainability and Insurance. The latter theme specifies technological dimensions that should be guaranteed: \textbf{open source} (of all software required to run the infrastructure), \textbf{open data} (of all relevant data necessary to replicate it), \textbf{available data} (i.e., the availability of underlying data as periodic data dumps) and \textbf{patent non-assertion} (i.e., avoid using patents to prevent the community to replicate an infrastructure). If followed strictly, POSI should guarantee the long term sustainability of infrastructures that provide open scholarly data and open source software that can be used to build service new and innovative services. Several infrastructures (including OpenCitations, Crossref and DataCite) have run self-assessment exercises to measure their compliance with POSI, as introduced in the POSI website at \url{https://openscholarlyinfrastructure.org/posse}.

\subsubsection{Data Collection}

The first layer of HuCI's architecture is constituted by a network of GLAM institutions playing an active role in the production and curation of citation data. While each institution is responsible for the extraction of citation data from their digitised holdings, we envisage the development of a common open source platform that can ease the tasks of extracting citations from publications as well as the manual curation of such citations. 

An example of software that could be deployed in the data collection layer is the Scholar Library (SL) platform\footnote{\url{https://github.com/ScholarIndex/ScholarLibrary}.} \cite{colavizza_linked_2018}. While it provides the typical functionalities of any digital library software (e.g., display of image and OCR), SL integrates specific components that perform the extraction of bibliographic references from digitized publications, and their disambiguation against bibliographic databases. In particular, it includes two components for the enrichment of publications with citation data: a machine learning-based citation extractor as well as a component to match bibliographic references against the unified catalogue of Italian libraries \cite{colavizza_references_2018}. 

The SL was designed to be \textit{deployed locally} while staying \textit{connected globally}: the local deployment ensures that digitized materials that cannot be shared openly remain private; APIs allow to harvest citation data from each local instance of SL and to connect them into a global citation index. As such, this platform could be deployed by partner institutions to facilitate the extraction and sharing of open citation data from their digitised holdings.


\subsubsection{Data Federation}

As a long-term solution to devising a scalable infrastructure for the storage of A\&H’s citation data we propose the HuCI virtual triple store, a federation of decentralised citation databases that can cooperate with each other by means of Web technologies, in particular RDF. Along the aforementioned lines, the interlinked databases of open citation data mentioned before have been recently released in order to address this aspect. The idea, in this aspect, is to organise existing and future open citations and scholarly metadata repositories (e.g., OpenCitations’ datasets, Wikidata, OpenAIRE) as part of a bigger and interlinked graph of open repositories\footnote{Something strongly supported by the 2017 report of the Confederation of Open Access Repositories (COAR), available at \url{https://www.coar-repositories.org/files/NGR-Final-Formatted-Report-cc.pdf}.}, which would allow them to scale in terms of their infrastructure and the amount of data they need to handle. This can be implemented by means of appropriate Web and Semantic Web technologies, such as RDF triplestores, which natively are able to handle federation in storing the data. The use of such technologies is also crucial for enabling the development of a decentralised network of interoperable Linked Open Data (LOD), which are hosted in several places. In particular, such interoperability should be guaranteed by using the same data model for exposing the citation data involved, as introduced below. In essence, HuCI is a virtual database, since it must be implemented as a federation of repositories which provide access to their citation data via the HTTP protocol according to a particular shared data model, and which enable to expose the data in multiple formats (CSV, JSON, RDF-based) to foster maximum understandability for both humans and machines. 

The feasibility of handling multiple and decentralised repositories of citation data effectively should be guaranteed by adopting a general metadata model in which the citation data will be described. If a different data model will be used by one of the repositories in the federated system, an explicit alignment to such general metadata model must be provided so as to make the federation possible.

Among the possible candidate data models for describing citation data there is the OpenCitations Data Model (OCDM) \cite{peroni_opencitations_2018}. OCDM is fully based on the Semantic Publishing and Referencing (SPAR) ontologies \cite{rutkowski_spar_2018} and other standard vocabularies (FOAF, PROV, etc.) for the specification of additional information about agents and provenance data. The data model is implemented by means of the OpenCitations Ontology (OCO)\footnote{\url{https://w3id.org/oc/ontology}.}, which is not yet another bibliographic ontology, but rather simply a mechanism for grouping together existing complementary ontological entities from several other ontologies, for the purpose of providing descriptive metadata all in one place. As introduced in \cite{pan_opencitations_2020}, the OCDM has already been adopted by several projects in the scholarly domain for organising bibliographic information such as the Venice Scholar Index\footnote{\url{https://venicescholar.dhlab.epfl.ch}.} \cite{colavizza_linked_2018}, the Linked Open Citations Database (LOC-DB)\footnote{\url{https://locdb.bib.uni-mannheim.de}.} \cite{lauscher_linked_2018} and the EXCITE Project\footnote{\url{http://excite.west.uni-koblenz.de}.} \cite{hosseini_excite_2019}.

\subsubsection{Service Layer}

In addition to the two layers dedicated to data collection and federation, HuCI will comprise a layer of services (e.g. Web APIs) that will enable and regulate the flow of data between HuCI, its network of data providers, external providers of bibliographic metadata, and providers of services for the enrichment of citation data (both internal and external). In particular, we envisage three types of services:
\begin{enumerate}
    \item services to harvest citation data from the network of data providers; 
    \item services to provide standardised access to external resources (e.g., archive and library catalogues);
    \item services to enrich the aggregated citation data (e.g., interlinking, deduplication).
\end{enumerate}

To the first type of services belong the APIs that will allow participating institutions to share the citation data extracted from their digitized collections. Citation data will be exposed by using the shared data model (such as the OpenCitations Data Model discussed above) and harvested via either SPARQL-based APIs or common Web REST APIs acting as a proxy to a SPARQL endpoint -- that can be easily set up using software such as RAMOSE \cite{daquino_creating_2020}, BASIL \cite{daga_basil_nodate}, grlc \cite{sack_grlc_2016}, OBA \cite{pan_oba_2020}, and SPARQL.anything \cite{daga_facade-x_2021}. These APIs could be available as part of HuCI or be offered by external providers. Provenance information, which includes the identification of the attribution, sources, activities and additional change tracking data, is also attached to the related citation data in order to allow trackability and restorability of citation data due to some, even unpredictable, changes \cite{peroni_document-inspired_2016}.

The second type of services aims to provide unified and standardised access to bibliographic metadata present in external resources, such as archive and library catalogues. These resources can be extremely valuable in various steps of the citation mining process (citation linking, author disambiguation), yet the heterogeneity of formats in which they are exposed hampers their reuse (see Section \ref{sec:metadata-ecosystem-requirements}). These services will facilitate the access to external resources by defining a common API specification for data exchange, as well as a common data format towards which individual bibliographic formats can be mapped.

Finally, a third type of services will provide enrichment of citation data, especially through interlinking and deduplication. In fact, due to the federated nature of HuCI, it may happen that the same bibliographic entity and its citations are stored multiple times in different repositories. Thus, it is crucial to provide mechanisms and algorithms for dealing with deduplication appropriately, both for live access to data for \textit{streaming} purposes, using a particular entry-point (e.g., a certain SPARQL endpoint), and to download full \textit{dumps} of citation data available in different federated repositories.

The resolution of these conflicts could be handled by using persistent identification schemas (like DOI, Handle, ORCID or VIAF) for uniquely identifying the various resources, or by applying disambiguation mechanisms based on entities’ metadata. Of course, the more persistent identifiers are specified for a bibliographic entity, the easier its disambiguation will be and, consequently, the deduplication of bibliographic resources coming from different repositories.

A good example of integrating remote services into a common research infrastructure comes from the recent project Open Mining INfrastructure for TExt and Data (OpenMinTeD)\footnote{\url{http://openminted.eu}.}. Their API specification for processing Web services defines a protocol that allows remote NLP components to be seamlessly integrated into processing pipelines (see e.g., \cite{ba_interoperability_2016}). Similarly, an API specification will need to be developed for external services that can be used to enrich HuCI's citation data. 

\subsubsection{Application Layer}

A crucial aspect of creating a citation index for the A\&H concerns the development of user interfaces allowing researchers to explore and exploit citation data. In the technical infrastructure we propose, search and visualization tools for the citation index will plug directly into HuCI’s virtual triples store and will constitute its application layer. 
This layer will comprise user interfaces for search and visualization, as well as software components that are meant to facilitate access to citation data stored in HuCI's virtual triples store via SPARQL API or via REST APIs built upon SPARQL endpoints. 

Several tools have been developed to date to display, analyze and visualize citation data. They differ substantially with respect to the platform where they run (Web or desktop), their main purpose (analysis, visualization, search), as well as the data sources for which they offer support (e.g., Web of Science, Scopus, PubMed, Microsoft Academics, OpenCitations, etc.). In particular, among these tools, there are VOSviewer \cite{van_eck_software_2010}, Sci2 \cite{sci2_team_sci2_2009}, CiteSpace \cite{chen_searching_2004}, Cytoscape \cite{shannon_cytoscape:_2003}, Bibliographic EXplorer (BEX) \cite{di_iorio_exploring_2015}, CiteWiz \cite{elmqvist_citewiz_2007}, Docudipity \cite{poggi_exploiting_2019}, CRExplorer (CREx) \cite{thor_introducing_2016}, Science Citation Knowledge Extractor (SCKE) \cite{lent_science_2018}, Scholia \cite{nielsen_scholia_2017}, the Scholar Index (SI) \cite{colavizza_linked_2018}, OSCAR \cite{heibi_enabling_2019}, and LUCINDA\footnote{\url{https://github.com/opencitations/lucinda}.}.

\section{Conclusions} \label{sec:conclusions}

In this article we have listed the main aspects that are necessary to devise the creation of a Humanities Citation Index (HuCI). We propose HuCI to be a decentralised and federated research infrastructure for gathering, sharing, elaborating, exposing bibliographic metadata and citation data of Humanities publications, that offers hooks for the development of further applications to keep track of the evolution of the Humanities research. 

The technical guidelines we have provided for the creation of such an infrastructure follows current trends shared by the Open Science community around the globe. Several of the principles regarding data sharing we proposed are grounded in existing guidelines such as the FAIR (findability, accessibility, interoperability, and reusability) data principles \cite{wilkinson_fair_2016}, which are considered a common and shared good practice in the field -- where the word \emph{data} in this context is an umbrella term including research data spreadsheets, software, workflows, slides, and other research objects that accompany a \emph{traditional} publication (e.g., a book, a journal article, a conference paper).

Several guidelines for enabling the creation of new open infrastructures -- including their technological compliance, plans for their long-term sustainability and governance -- have been proposed in the past five years, and have directly guided our work on HuCI. The Principles for Open Scholarly Infrastructures \cite{bilder_principles_2020}, the work done by the Confederation Of Open Access Repositories (COAR) on best practices for implementing digital repositories \cite{coar_wg_next_generation_repositories_behaviours_2017, confederation_of_open_access_repositories_coar_2020}, and other principles proposed by independent scholars such as the TRUST (transparency, responsibility, user focus, sustainability and technology) principles \cite{lin_trust_2020} have been extensively reused and adapted to devise the various component of the technical research infrastructure in HuCI. The very same principles characterise several national and international initiatives, such as the community workshop held in 2021 with the aim of shaping the main technical and organisational aspects for the creation of a open knowledge base of scholarly information for the Netherlands \cite{neylon_open_2021}, and organisations created to help open infrastructures flourishing, such as the Global Sustainability Coalition for Open Science Services (SCOSS)\footnote{\url{https://scoss.org}.} and Invest in Open Infrastructures (IOI)\footnote{\url{https://investinopen.org}.}.

As part of our future work towards the creation of HuCI, we plan to conduct a survey among Humanities scholars in order to elicit their views and desiderata with respect to the prospects and usefulness of such a citation index. This survey could be conducted in coordination with ongoing international activities on the topic of bibliographic data in the Humanities, notably the DARIAH-EU Bibliographic Data Working Group\footnote{\url{https://www.dariah.eu/activities/working-groups/bibliographical-data-bibliodata}.}. Nevertheless, given the striking similarities that citation indexes bear with thematic bibliographies (both printed and digital) -- which are widely used by scholars across the Humanities -- it does seem plausible to postulate that such a citation index will meet the interests of many scholars.

Our hope is that the guidelines, principles, and technological approaches described in this work can be an appropriate starting point for the implementation of HuCI, a fundamental tool for Humanities research. The goals depicted by HuCI, and their technical implementation, are possible only if the Humanities scholars and institutions act together in a decentralised and coordinated fashion, by sharing efforts, resources, and services towards a common objective, of which the suggestions in this article represent only the starting point. 

\backmatter

\bmhead{Acknowledgments}

We thank the anonymous reviewers for their constructive feedback. The work of Matteo Romanello has been supported by the Swiss National Science Foundation under grant number PZ00P1\_186033. The work of Silvio Peroni has been partially funded by the European Union's Horizon 2020 research and innovation program under grant agreement No 101017452.

\bibliography{bib/HuCI-zotero}
\end{document}